\documentclass{elsart}

\usepackage{psfig}

\begin{document}

\begin{frontmatter}
\title{
Curious properties of the recycled pulsars
and the potential of high precision timing.
} 
\author{Matthew Bailes}
\address{
Centre for Astrophysics and Supercomputing, Swinburne 
University of Technology, Mail 39, PO Box 218, Hawthorn, Vic, 3122, 
Australia. mbailes@swin.edu.au
}

\begin{abstract}

Binary and Millisecond pulsars have a great deal to teach us about
stellar evolution and are invaluable tools for tests of relativistic
theories of gravity. Our understanding of these objects has been
transformed by large-scale surveys that have uncovered a great deal of
new objects, exquisitely timed by ever-improving instrumentation.
Here we argue that there exists a fundamental relation between the spin
period of a pulsar and its companion mass, and that this determines
many of the observable properties of a binary pulsar. No recycled
pulsars exist in which the minimum companion mass exceeds (P/10 ms)
M$_\odot$. Furthermore, the three fastest disk millisecond pulsars
are either single, or possess extremely low-mass companions ($M_{\rm
c} \sim 0.02$M$_\odot$), consistent with this relation. 
Finally, the
four relativistic binaries for which we have actual measurements of
neutron star masses, suggest that not only are their spin periods
related to the companion neutron star mass, but that the kick
imparted to the system depends upon it too, leading to a correlation
between orbital eccentricity and spin period.  
The isolation of the relativistic binary pulsars in the magnetic
field-period diagram is used to argue that this must be because the
kicks imparted to proto-relativistic systems are usually small,
leading to very few if any isolated runaway mildly-recycled pulsars.
This calls into question the magnitude of supernova
kicks in close binaries, which have been usually assumed to be
similar to those imparted to the bulk of the pulsar population.
Finally, we review some of the highlights of the Parkes
precision timing efforts, which suggest 10 nanosecond timing is
obtainable on PSR J1909--3744 that will aid us in searching
for cosmological sources of gravitational waves.

\end{abstract}

\end{frontmatter}

\section{Introduction}

The field of radio pulsar astronomy was born with the discovery of the
first pulsar by Bell and Hewish \cite{hbp+68}. Early surveys with the
world's largest radio telescopes soon swelled this number to in excess
of 350 by the early 1980s\cite{lmt85}. Although these surveys were
very successful at finding slow, high-field pulsars, the fraction of
known pulsars in binaries at the time was extremely small ($\sim$1\%).  The
discovery of the first binary pulsar\cite{ht75a} had a profound impact
on both astronomy and fundamental physics. It was soon realised that a
binary pulsar could be used in tests of relativistic gravity
\cite{bo75}, but also to constrain the nature of
supernovae\cite{sut77}, the masses of neutron stars\cite{ht75a}, and
how binaries evolve and die\cite{fv75}.

In their classic paper, Bhattacharya and van den Heuvel (1991)
\cite{bv91} documented the evolutionary scenarios that explained the
expanding binary pulsar zoo. Inspired by PSR B1913+16, van den Heuvel
elegantly explained the nature of binary pulsar and others
such as PSR B0655+64 and PSR B0820+02 as the result of a
binary evolution in a series of papers e.g.\cite{sv82,vdh84}.  This
involved several radical ideas, including binary mass transfer,
common-envelope evolution, pulsar recycling, supernova kicks, tidal
circularization, the widening of pulsar orbits via mass
transfer, and accretion-induced collapse (AIC) of white dwarfs.
Pulsars with low-mass companions were explained by either
recycling or the accretion-induced collapse of a white dwarf.
Bailes (1989)\cite{bai89} argued that it was possible to create all the 
recycled pulsars from recycling of neutron stars if field decay
was caused by accretion, without the need for AIC, 
and suggestions for how this might occur have been proposed
in the literature by various authors\cite{rom90,sbmt90}.

Hobbs et al.\cite{hllk05} and Brisken et al.\cite{bbgt02} have
conclusively demonstrated that most isolated pulsars possess
large velocities consistent with a large kick at birth. If all pulsars
receive such kicks, then we expect general misalignment of pulsar spin
and orbital angular momentum axes in binary pulsars, as well as highly
eccentric orbits\cite{hil83,cb05}. Recently, Dewi et
al. (2005)\cite{dpp05} have called into question the ability of a
supernova from a close binary to deliver a significant ($V>100$ km
s$^{-1}$) kick to a proto-neutron star.  Dewi et al.'s work
demonstrates that some simple assumptions can reproduce the rough
correlation observed between spin period and orbital eccentricity
noted by Faulkner et al. (2005)\cite{fkl+05}. Their work assumed that
the spin period was determined by the duration of the mass accretion
phase, with a timescale set by the mass of the donor. Low-mass donors
could spin up pulsars to faster spin periods. Although they did not
investigate the millisecond ($P<10$ ms) pulsars with white dwarf or
low-mass companions, their work on double neutron stars showed
how such a correlation was a logical consequence of small kicks and
spin periods that were related to the duration of the mass tranfer
phase.

Pulsar astronomers have been continually improving their
instrumentation and timing methodologies in order to measure
quantities with astounding precision\cite{tfm79,ktr94,vbb+01} to
reveal more and more subtle effects that allow physical constraints to
be derived about their binary hosts. These measurements, and the
growing millisecond pulsar (MSP) population have helped theorists develop
models to understand binary evolution and the nature of supernovae.
The longevity of Bhattacharya and van den Heuvel is testament to
the fundamentals inherent in the paper. Since its publication however,
new types of binary pulsar have emerged.  Whilst some of these have
proven to be dramatic confirmations of the models, others require us
to improve on the recipes summarised by Bhattacharya and van den
Heuvel (1991).

In this paper I will review the binary and recycled pulsar population,
examine some interesting correlations and trends, and suggest modifications
to models that the new data appear to demand. I also demonstrate
how the highest precision pulsar timing is capable of delivering arrival
times down to accuracies of near 10 nanoseconds if we can eliminate
sources of systematic error in pulsar timing.


\section{The Binary Pulsars}

\subsection{General Characteristics}

One of the most popular diagrams for understanding the origin
and evolution of binary pulsars is the Magnetic field-Period 
($B$-$P$) diagram shown in Fig 1. Although unpopular
with some observers, who usually substitute the observed period
derivative for the theorist's ``magnetic field'', $B^2 \propto P \dot P$,
it shows that the binary pulsars have much weaker fields
and shorter spin periods than the majority of the pulsar
population. The recycling model explains their location via
the ``spin-up'' of an otherwise normal high-field neutron
star and the destruction of its field in the accretion
process. This is often referred to as ``recycling''.

\begin{figure}[ht]
\centerline{\psfig{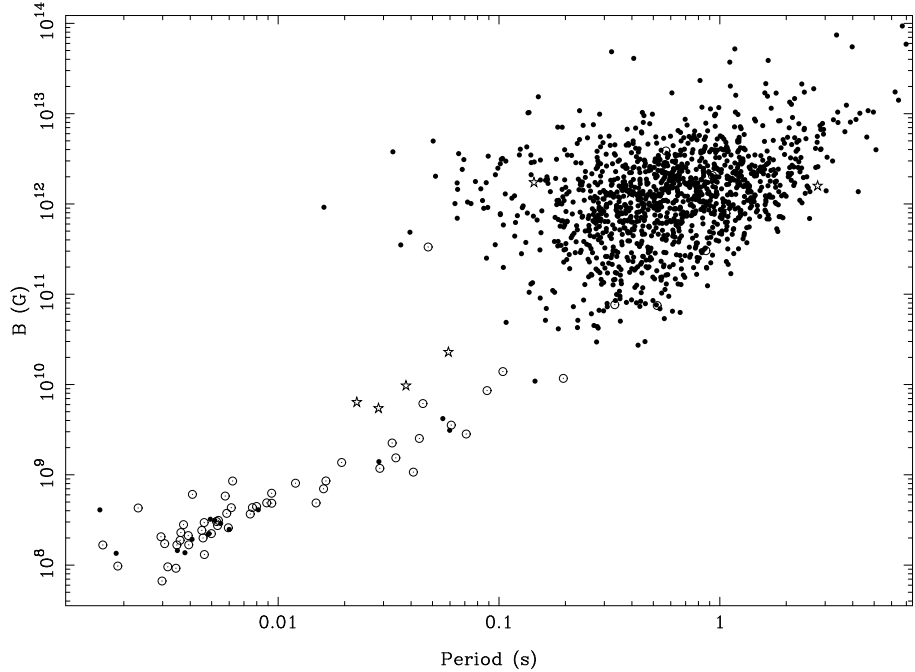}}
\caption{
The pulsar ``$B$-$P$'' diagram for all pulsars not associated with
globular clusters. Data are taken from the ATNF pulsar catalogue
with additions from \cite{jhb+05}, \cite{fkl+05},
\cite{sfl+05}, \cite{bjd+06}. Binary pulsars are indicated by circles with
dots inside them. The relativistic binary pulsars are represented
by stars and we define these by having an orbital
period less than a day. Note that the double pulsar appears twice in this
diagram. Improvements in pulsar search instrumentation have
only made the possibility of large numbers of pulsars in the
lower left of the diagram since about 1990.}
\label{fg:bp}
\end{figure}

\subsection{Spin Period-Companion Mass Correlation}

In simple terms, the recycling model for binary pulsar
evolution predicts that recycled pulsars should have
spin periods that are related to their mass accretion
history, and magnetic field strength.
Low-mass stars evolve more slowly than high-mass stars, and
have more time to sustain mass accretion. Thus, one of the predictions
of the recycling model is that there should be some relation
between spin period and companion mass soon after mass accretion
ceases, but only if the magnetic field does not limit the
accretion through the action of a propellor phase. 
To explore this, in Fig 2 we plot the minimum
companion mass of a binary pulsar (assumes an edge-on
orbit and a pulsar $>1.35$ M$_\odot$) against the currently
observed spin period which shows good agreement with the
recycling model. 
The only serious competing model for
millisecond pulsar formation, that of accretion-induced 
collapse of a white dwarf\cite{bg90} to directly produce a millisecond
pulsar should have no such correlation.

\begin{figure}[ht]
\centerline{\psfig{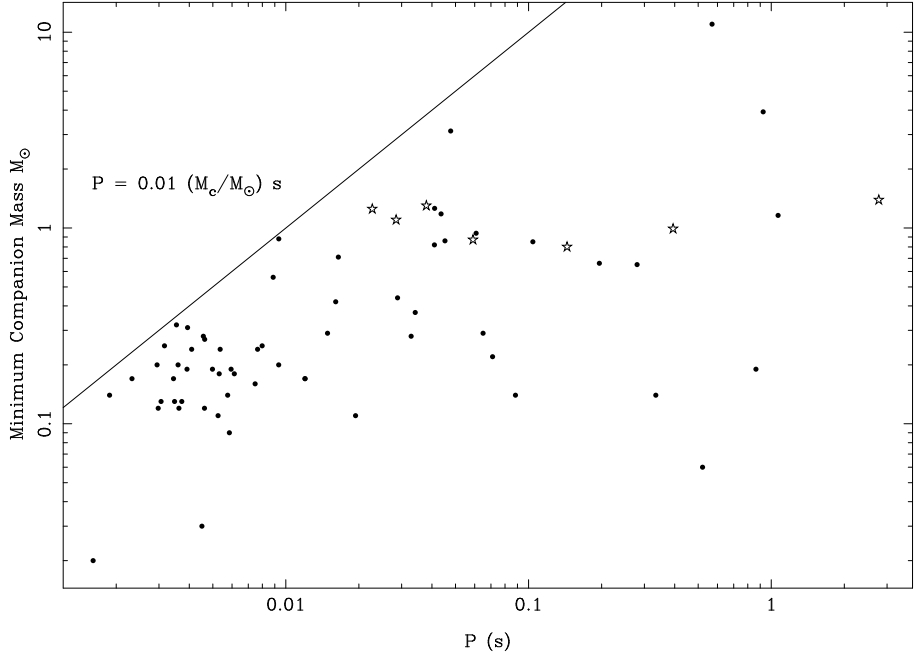}}
\caption{
The minimum companion masses of all the field binary
pulsar population plotted against their spin periods.
A striking linear barrier is revealed with no short-period
pulsars possessing massive companions. 
Pulsars with stars as their symbols have eccentric 
orbits ($e>0.05$) and tend to be more massive. We understand
this as their companions have exploded, and should have
left behind a stellar remnant near the Chandrasehkar mass
of 1.35 M$_\odot$.
}
\end{figure}

From Fig 2 we can determine the handy phenomological
relation:
\begin{equation}
M_{\rm c} < P / (10 {\rm ms}) {\rm M}_\odot
\end{equation}
\noindent 
where $M_{\rm c}$ is the minimum mass of the companion.

Although we might have expected some rough relation
between spin period and mass, or perhaps no $\sim$1 ms
pulsars to possess heavy companions,
what is striking is that equation 1
appears to define an exclusion region that is in effect
for {\it all} recycled pulsars.
The three fastest ($P<$2\,ms) millisecond pulsars
in the field are either solitary (PSR B1937+21, PSR J1843--1113)
or have extremely low-mass ($M_{\rm c} \sim 0.02\, $M$_\odot$) 
companions (PSR B1957+20). This means that it is the companion mass,
not exotic gamma-ray death-rays or the like that determine which pulsars
remove or almost destroy their companion stars. 
The conclusion is
that to be a rapidly spinning ($P<2$ ms) pulsar, the companion must
ultimately be of very low ($M_{\rm c}< 0.2\,$M$_\odot$) mass or completely
destroyed. We might be tempted to conclude that all
of the solitary millisecond pulsars with periods
less than 10\,ms were once spinning more rapidly and
had low-mass companions. Certainly, if it were the magnetic
field of the pulsar, and not the companion mass of the
pulsar that determined the final spin period, we might not
expect any such relation to hold between the spin period
and companion mass. 

\subsection{Companion Mass of Significantly Eccentric ($e>0.03$) Binaries}

A prediction of the recycling model is that in order to have a
pulsar appear in a binary with a significantly non-eccentric orbit the
companion must be an inert object, like a neutron star, white dwarf or
main sequence star well inside its Roche Lobe.
Although it is possible to produce a pulsar
in an eccentric binary with a very low mass main sequence star, none
have ever been observed.  This might be because these systems often
disrupt during the supernova explosion, are short-lived, or the
extreme initial mass ratio required is rare.

We now know of several binary pulsars orbiting massive companions like
PSR B1259--63 and PSR J0045--7319.  To find a pulsar in an eccentric
orbit about a white dwarf companion requires a progenitor in a system
with an initial mass ratio close to unity\cite{ts00}, which always
leaves behind a massive white dwarf such as in PSR B2303+46 or PSR
J1141--6545.  Finally, neutron stars are heavy objects, and appear to
have minimum masses near 1.25-1.4 M$_\odot$.  As shown in Fig 2, the
relativistic eccentric binaries are all consistent with reasonably massive
companion stars, as we might expect.

\subsection{Eccentricities of the low-$e$ ($e<<0.03$) Recycled Pulsars}

\begin{figure}[ht]
\begin{center} 
\begin{tabular}{c}
\mbox{\psfig{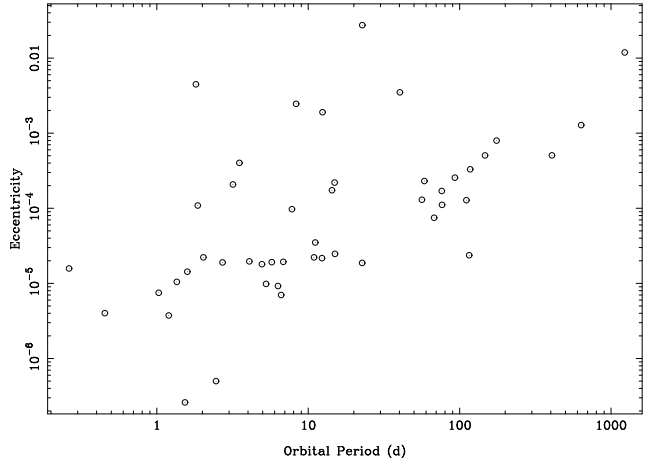}}
\end{tabular}
\caption{
Orbital eccentricity of all the binary pulsars with $e<0.03$
that we believe have a high probability of being white
dwarf companions due to their circularity and low mass.
The most eccentric MSP (PSR J1618--3919) has a spin
period of 11 ms and $e=0.027$. Only field, as opposed to
globular cluster pulsars are shown.
}
\end{center}
\end{figure}

It takes time to spin up a pulsar to millisecond periods
due to the amount of mass involved ($\sim 0.1\,$M$_\odot$) and the Eddington
limit for mass accretion. During this time, we might expect 
the orbit to become highly circular. Indeed, we would expect that
the shorter the spin period, and the tighter the orbit, the
more circular it might become. In Fig 3 we show the
orbital eccentricities (where measured) of all the binary
pulsars with $e<0.03$. There are many slower
period pulsars with $0.00004 < e < 0.03$ even at short
orbital periods and no clear trends. This is consistent with
the duration of the recycling phase playing a crucial role
in removing eccentricity. 

\begin{figure}[ht]
\begin{center} 
\begin{tabular}{c}
\mbox{\psfig{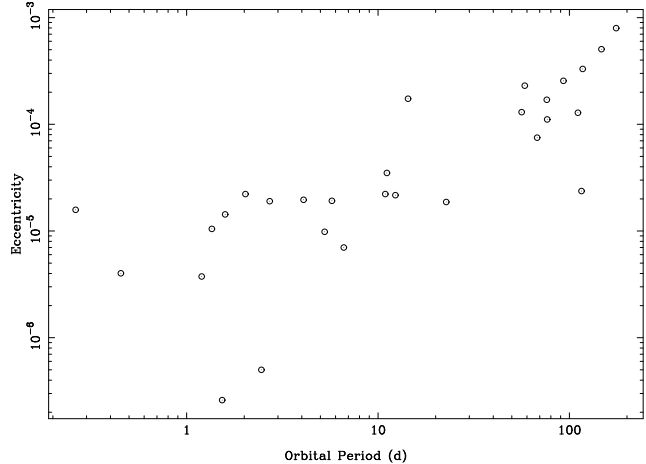}} \\
\end{tabular}
\caption{
Orbital eccentricity of all disk MSPs
with $P<10$ ms. Sustained recycling appears to make
most of the MSPs conform to a nice orbital eccentricity
orbital period relation. Pulsars with upper limits are
not shown.
}
\end{center}
\end{figure}

In Fig 4 only those pulsars with $P< 10$ ms are shown. A few features
are immediately apparent. Firstly, all binary millisecond
pulsars with $P<$10 ms and orbital periods ($P_{\rm b}$)
$<$10 d have eccentricities $<2.5\times10^{-5}$.
Secondly, although there is a trace of a nice trend for
pulsars with periods greater than 50 d to have a 
correlation of eccentricity and orbital period, 
Stairs et al. recently showed that PSR J1853+1303 has a small eccentricity
despite a period in excess of 100 days. Clearly,
long periods of mass accretion are not guaranteed to
iron out eccentricity, and long orbital period MSPs
can still possess very small orbital eccentricities.
PSR J1618--3919 is a bizarre object with a spin period
of 11 ms and an orbital eccentricity of 0.027 and is clearly anomolous. 
Finally, there is a dearth of MSPs with orbital
periods between 12 and 60\,d, as first noted by Camilo\cite{cam95}.

\subsection{The Binary Fraction of Recycled Pulsars}

\begin{figure}[ht]
\centerline{\psfig{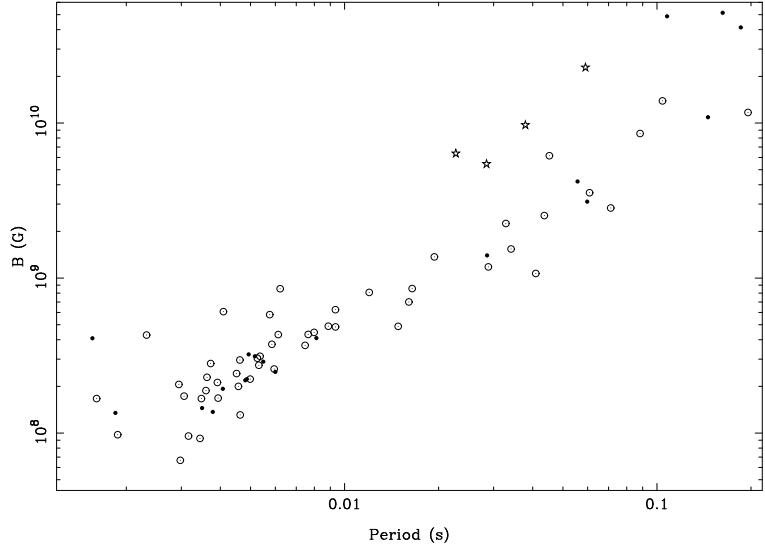}}
\caption{
The lower left of Fig 1 showing the paucity of solitary
millisecond pulsars (simple dots) between 8 and 60 ms,
and the relatively high magnetic fields of the relativistic
pulsars. If the solitary recycled pulsars with similar
periods to the relativistic binaries are just the result
of random kicks that would have otherwise kept PSR B1913+16-like
binaries together, it is hard to understand why their fields
are lower.
}
\end{figure}

In Fig 5 we show the lower left region of the $B$-$P$ diagram
where the recycled pulsars live. A few interesting points
can be established. Firstly, as previously discussed,
at short periods ($P<2$ ms) 
all of the pulsars are either solitary or have very low-mass
companions. Secondly, between 3 and 8 ms, the solitary and binary
pulsars largely
overlay each other, giving little clue as to what makes some
solitary and others binary. Between 8 and 60 ms the population
becomes almost 100\% binary, although a recent discovery 
(PSR J1038+0032) has now appeared in this gap\cite{bjd+06}.

Finally, the relativistic eccentric binaries, indicated by
stars in the Figure, have larger fields than comparative 
period circular-orbit systems and any isolated recycled pulsars.
Since they have larger mass
companions we might be tempted to suggest that this left
less time for mass accretion-induced accretion. Since the
four relativistic binaries (defined here by coalescence times
less than a Hubble time) are all above the band of circular
orbit and isolated systems, we conclude they had a different origin.
If the progenitor systems had had a high disruption probability 
during the final supernova, we would expect many isolated
pulsars with similar periods and $B$-fields to populate
the diagram. They do not. From this we conclude that
the disruption probability is small.

In the recycling model, radio pulsars achieve their high
velocities from asymmetric explosions that impart a large
velocity to the pulsar. If the probability of survival is
high, then there should be relatively few solitary pulsars
like the eccentric binaries for every binary pulsar. If, on
the other hand, the disruption probability is great, then
we would expect many solitary recycled pulsars for every 
eccentric binary. Classical mechanics tells us that
a pulsar's orbital eccentricity is a measure of how
near it came to becoming unbound in the final explosion.
PSR B1913+16's eccentricity of 0.617 suggests it came
close to disruption, whereas PSR J0737--3039A/B's $e$ of
just 0.088 suggests a quiet explosion and little or no
associated kick. 

\subsection{The Near-circularity of the Relativistic Binaries?}

One of the greater mysteries under current consideration
is why so many of the relativistic binaries have really
very circular orbits ($e<0.3$).
Chaurasia and Bailes (2005) \cite{cb05} postulated
that this might be because low-e systems have much longer
observable lifetimes than high-e ones, but this can only
be part of the story. In the very young binaries PSR J1141--6545
and PSR J1906+0746 this selection effect appears unimportant
and yet both have very small eccentricities (0.17 and 0.085 respectively).
Large random kicks would place the majority of systems into very 
eccentric orbits. Indeed it has been shown\cite{cb05} that if kicks are
comparable to the relative orbital velocity of the stars and fill
the available phase space by chance, then
only 10\% of the remaining bound systems should have $e<0.3$. Instead,
of the seven relativistic binaries with orbital periods less
than 2 days, an amazing 6 have $e<0.3$. The probability of
this is minute, and is very strong evidence that the sort of
kicks pulsars must receive to obtain their space velocities
of several hundred to 1000 km s$^{-1}$ are not commonplace in
close binaries. Indeed, even if there are kicks, to get low
eccentricities requires supernova masses only just above
the Chandrasehkar limit, as asserted for the double pulsar
by Dewi et al. (2005)\cite{dpp05}.

Until the discovery of the solitary $P=28$ms PSR J1038+0032 it would
have been tempting to suggest that we could always map the spin
period to the duration of the mass transfer phase, and
hence mass of the progenitor of the exploding star, and
the subsequent kick.
Longer-period recycled pulsars would then have had large
exploding companions, that received a large kick. This
then placed them in orbits so eccentric that we had
little time to observe them before they coalesced, left
some in PSR B1913+16-like orbits, and disrupted many others.
Stars that had longer accretion phases only did so because
their companions were less massive, and exploded when
the pre-supernova mass was only just above that of the
Chandrasekhar limit leaving behind a young pulsar that
received almost no kick. However, if PSR J1038+0032
is the result of a disrupted binary, then this is not
consistent with such a simple model.
To retain the model, it is then necessary to
explain the existence of PSR J1038+0032 
as a pulsar that destroyed its companion rather than
was disrupted in the final supernova of a
binary that produced two neutron stars.

In summary, in order to explain the
orbital eccentricities of the binary pulsars, 
the kicks imparted to the
last-born pulsars in a relativistic binary are small,
and the mass lost in the final supernova modest. 

\subsection{Another test of the spin-companion mass correlation}

How else might we test the hypothesis that the companion mass
determines the spin period of a neutron star?  In a relativistic
binary, we can not just establish a minimum companion mass, but
actually measure the companion mass.  If what we have seen about spin
periods being related to companion masses is true in the binary pulsar
population as a whole, we might expect the longest period recycled
pulsars to have the heaviest neutron star companions.

We are now fortunate enough to have four relativistic binaries where
the companion mass is not just a lower limit, but a measurement.  In
Fig 6 we plot the spin period of the four recycled relativistic
pulsars PSRs B1913+16, B1534+12, J0737--3039A, and J1756--2251 against
their companion neutron star mass.  Inspection of Fig 6 reveals the
anticipated correlation between pulsar mass and recycled pulsar period, 
providing further support for the
universality of equation 1.

\begin{figure}[ht]
\centerline{\psfig{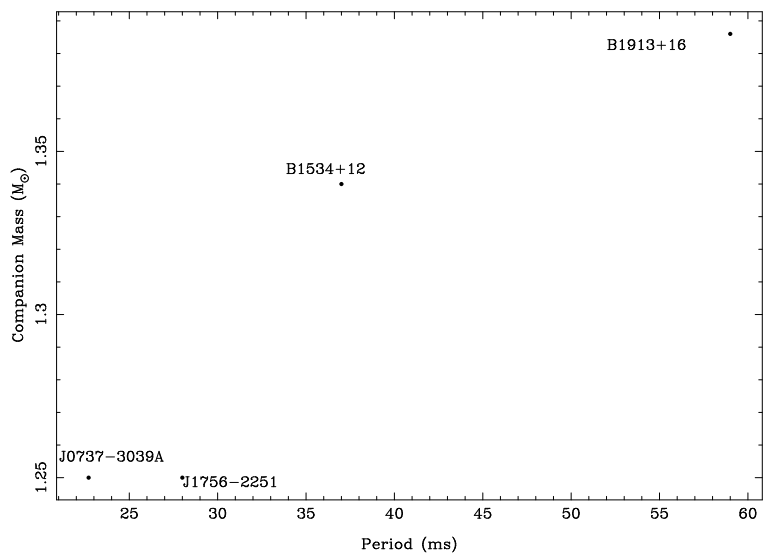}}
\caption{
Companion masses against spin period for the four relativistic
binary pulsars. PSR J1756--2251's mass is an upper limit.
}
\end{figure}

\section{Precision Pulsar Timing}

Millisecond pulsars have a stability which is comparable
to that of terrestrial atomic clocks. This makes them
ideal targets for long-term timing to investigate 
cosmological sources of gravitational radiation.
For many years now we have been undertaking
systematic surveys for millisecond pulsars suitable for
precision timing experiments with the Parkes 64\,m telescope.

To be a good pulsar for timing, several characteristics 
are desirable. Firstly, it must be possible to obtain
sub-$\mu$s residuals on a routine basis. This means
the average flux density should be high, and the
pulsar's pulse should have narrow features. Secondly,
the pulsar should be intrinsically stable, and not
exhibit random excursions from regular rotation, often
referred to as ``timing noise''. 
Finally, it helps if
the pulsar has some measureable quantity of astrophysical
interest like Shapiro delay, 
or is close enough to measure its distance through parallax
or some more subtle means.

Two pulsars stand out from the 50 or so discovered at Parkes
in the last 15 years. These are PSR J0437--4715, and PSR J1909--3744.

\subsection{PSR J0437--4715}

PSR J0437--4715 is an otherwise unremarkable pulsar that
is only $\sim$ 150 pc from the Sun. This proximity to the
Earth makes it very special, making its humble luminosity
of $\sim$3 mJy kpc$^2$ at 20 cm result in a mean flux of
100 mJy at 20cm, two orders of magnitude above most of the
MSPs known. 

Early efforts on this pulsar reported its basic parameters
\cite{jlh+93}. Soon the proper motion became apparent
in the timing\cite{bbm+95} and more and more subtle effects
have become measureable with new instruments.
van Straten et al.\cite{vbb+01} report the most precise
timing of this pulsar with an RMS residual of just 130 ns.
Recently \cite{hbo06} have extended the timing of this
pulsar to over ten years. Almost a decade ago, it was
realised that the orbital period derivative of this
pulsar would be not only measureable, but lead to
an independent measure of the distance via the Shklovskii
effect\cite{bb96}. For PSR J0437--4715, we
now have a reliable measure of the period derivative and
find that the implied distance is 167(1) pc. The error is the formal
error from the timing model, and does not take into account
any pulsar timing noise or long-term period wandering that
might affect the orbital period derivative. The size of the
error is comparable to the relative contribution of 
acceleration towards the Galaxy compared to the Shklovskii
term. This result is being prepared for publication and
the systematic errors investigated (Verbiest
et al. 2006). 

\subsection{ PSR J1909--3744}

Discovered in Jacoby's extension of the Swinburne
Intermediate Latitude survey, PSR J1909--3744 has
an extremely narrow pulse profile of just 42 $\mu$s
and arrival times of astonishing accuracy. Timing
of this pulsar has revealed that the companion is
a well-determined 0.202(1) M$_\odot$, and the pulsar 
1.44(3) M$_\odot$ \cite{jhb+05}. The DM of only
10.34 pc cm$^{-3}$ causes the flux of this pulsar
to vary by over a factor of 30 on timescales of
hours. At scintillation maxima, the arrival times
are of phenomenal accuracy, and in theory well below
100 ns.

In pulsar timing, systematic errors are often the
limiting factor. PSR J1909--3744's
narrow pulse make it relatively immune to the disturbances
in the apparent pulse shape exhibited by other pulsars
from imperfect receivers and instrumentation.
To try and establish if we could time pulsars at
much less than 100ns accuracies, over a three week
period in 2004 we observed PSR J1909--3744 almost
every day for 3 weeks, trying to ensure one scintillation
maxima per day. The resulting arrival times
are shown in Fig 7. Now that we have extended the
time baseline on this pulsar beyond 3 years, we can
eliminate these three weeks of data from our fit, and
see what RMS residual we can obtain if we just take
the 3 weeks of data and only fit for a DC offset.
We find the 121 data points give an RMS of 122 ns.
This means that we can possibly time this pulsar
over any given 3 week period to an accuracy of 11 ns.

\begin{figure}[ht]
\begin{center} 
\begin{tabular}{c}
\mbox{\psfig{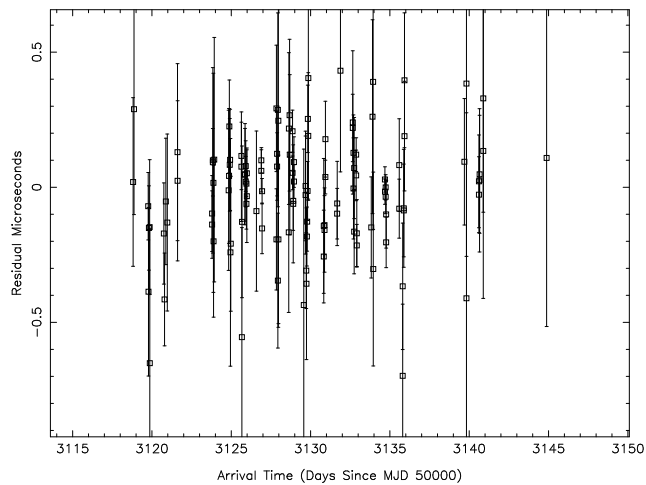}}
\end{tabular}
\caption{
Timing of PSR J1909--3744 over a 3 week period. The 121
arrival times give an RMS residual of only 122 ns, which
has a standard error of the mean of just 11 ns. Arrival times
of this accuracy have tremendous importance in the search
for a gravitational wave ``background''.
}
\end{center}
\end{figure}

By observing this source with larger bandwidths, cooler
receivers or just more often, we could potentially
get this sort of arrival time accuracy in just a few
days. Sensitivity to a cosmological gravitational
wave background will require more pulsars like PSR J1909--3744
for success.

\section{Conclusions}

The observed properties of the recycled pulsars tell us
that the final companion mass largely dictates the spin
properties of pulsars.

There are some important consequences:
\begin{itemize}
\item
The relativistic pulsars with the tightest orbits should
have low runaway velocities, and spend most of their
observable lifetimes near the Galactic plane.
\item
To obtain several hundred Hz spins, pulsars require
a low-mass $M_{\rm c}<0.3$ M$_\odot$ companion.
\item
Close binaries that produce relativistic binaries
eject little mass in their final
supernova explosion, and have small kicks,
subsequently there are few runaway isolated recycled pulsars.
\item
Neutron star companion masses of relativistic binaries
are related to the spin period of the recycled pulsar.
\end{itemize}

Long-term precision timing can yield both extremely
accurate distances and arrival times of astounding
precision suitable for detecting a gravitational
wave background.

\section*{Acknowledgements}
I wish to thank Professor Ed van den Heuvel for his
brilliant illustration of binary evolution with his
beautiful cartoons that helped me enter this field, 
for his generous refereeing of my
Ph D thesis and some early papers, his friendship,
and production of an army of disciples with which
to discuss interesting ideas about everything to
do with binary pulsars. Joris Verbiest aided me
in preparation of the 0437 data shown at the conference.


\end{document}